\documentclass[aps,pra,showpacs,twocolumn,superscriptaddress]{revtex4-1}

\usepackage{amsmath}
\usepackage{amsfonts,bbold}
\usepackage{amssymb}
\usepackage{graphicx}
\usepackage{subfigure}
\usepackage{natbib}
\usepackage[colorlinks=true,linkcolor=blue,urlcolor=black,citecolor=blue]{hyperref}

\newcommand{\tr}{\mathrm{Tr}}

\newcommand{\eqapxrwa}{(B4)}
\newcommand{\eqapxsz}{(C2)}

\begin{document}

\title{Memory-induced geometric phase in non-Markovian open systems}

\author{Da-Wei Luo}
\affiliation{Beijing Computational Science Research Center, Beijing 100094, China}
\affiliation{Center for Quantum Science and Engineering,  and Department of Physics, Stevens Institute of Technology, Hoboken, New Jersey 07030, USA}

\author{J. Q. You}
\affiliation{Beijing Computational Science Research Center, Beijing 100094, China}

\author{Hai-Qing Lin}
\affiliation{Beijing Computational Science Research Center, Beijing 100094, China}

\author{Lian-Ao Wu}
\affiliation{Department of Theoretical Physics and History of Science, The Basque Country University (UPV/EHU), PO Box 644, 48080 Bilbao, Spain}
\affiliation{Ikerbasque, Basque Foundation for Science, 48011 Bilbao, Spain}

\author{Ting Yu}
\email{Corresponding authors: T.Y. ting.yu@stevens.edu}
\affiliation{Center for Quantum Science and Engineering,  and Department of Physics, Stevens Institute of Technology, Hoboken, New Jersey 07030, USA}

\date{July 17, 2018}

\begin{abstract}
Geometric phases have been shown to be feasible in implementing quantum gates to perform quantum information processing. For all the realistic applications, the environmental influence on
the geometric phase and decoherence such as memory effects must be properly considered in order to achieve the required precision in geometric quantum computation. In this paper, we study the geometric phase for a generic open quantum system based on a microscopic model. A remarkable feature of the open system's geometric phase obtained from our theoretical formulation is that the geometric phase can be obtained regardless of the existence of the master equations, while the environmental noise features such as memory effects are fully accounted for. We demonstrate that the geometric phases for a general open quantum system can be fundamentally modified by its non-Markovian environments.  

\end{abstract}


\maketitle


\section{Introduction}
As a quantum state slowly evolves, it will acquire a global phase, which contains a geometric contribution that is only dependent on the path transversed by the system. 
This phenomenon was extensively studied by Berry in a seminal paper~\cite{Berry1984} and was later generalized to non-adiabatic evolutions~\cite{Aharonov1987} and non-cyclic evolutions~\cite{Samuel1988,Pati1995}. Over the last decades, the geometric phase has been studied in various situations such as nuclear resonance \cite{Tycko1987a}, quantum Hall  effects~\cite{McCann2006} and quantum phase transitions~\cite{Carollo2005}. Recent growing interest in quantum information and quantum computing~\cite{Nielsen2000} has contributed  to a crucial realization that the geometric phase can be an important resource in implementing quantum gates in geometric quantum computation. One of the major advantages of realizing universal quantum computation with geometric phase is its intrinsic resilience against errors and perturbations~\cite{Zanardi1999}. 

 In quantum information processing (QIP), however,  the quantum systems used as information carriers are inevitably coupled to the surrounding environments causing decoherence and dissipation, hence,  any realistic applications in QIP must take the open-system effects into account.  While the geometric phase for a quantum open system can be studied by using either the Markov quantum jump approach~\cite{Carollo2003}, Markov quantum state diffusion (QSD) equations~\cite{Buri2009,Bassi2006} or by using quantum state purification approach ~\cite{Tong2004}, the geometric phase for a  general quantum system has still not been properly addressed due to the lack of efficient analytical approach to recovering density operator and the mathematical complexity 
 arising from the environmental memory effects. In this paper,  we report a generic approach to the geometric phase of quantum open systems through the non-Markovian diffusive trajectories that unravel the open system dynamics~\cite{Diosi1998,Yuetal1999,Jing2010}. Our approach directly explores the geometric phase on a single quantum process level and fully accounts for the memory effect of the environment without using the explicit master equations, which are typically not available for general open systems.
  
More precisely, we will consider the geometric phase for quantum trajectories generated by the exact stochastic Schr\"{o}dinger called non-Markovian QSD equation, which describes non-unitary and non-cyclic evolutions without any approximations. The ensemble mean of generated quantum trajectories  will crucially recover the density matrix of the open system under consideration.  Notably, the geometric phase of a generic open system can be computed without the information of the (exact) master equations that are typically difficult to obtain except for a few special cases including the Born-Markov approximation \cite{Breuer2002,Carmichael,Knight}.  Our general formalism is exemplified with two exactly solvable models exhibiting several new features of the geometric phase when the memory effect caused by the environment cannot be ignored. We emphasize that our approach to the open system geometric phase is generic and expands our understanding of the power of quantum geometric phase in a new domain where the environment can be much more realistic. In the light of quantum computing application, besides the numerical advantage in tracking real-time information of the geometric phase, our formalism has recovered an important geometric phase correction due to the memory effects that has been missing in all the standard Born-Markov approximations \cite{Knight,Carollo2003}.

\section{Geometric phase in an open quantum system: general definition}

Consider a generic quantum system described by $H_{\rm sys}$ that is embedded in a multi-mode bosonic bath, which is generally referred as an environment in this paper,  and the total system consisting of the system of interest and its environment initially prepared in a pure product state $|\Psi (0)\rangle=|\psi_{\rm sys}(0)\rangle\otimes |\varphi_{\rm env} (0)\rangle$ where the environment initial state is a vacuum $|\varphi_{\rm env} (0)\rangle=|0\rangle$. The Hamiltonian of the total system is given by
\begin{equation}
\label{total_H}
    H_{\rm tot}=H_{\rm sys}+\sum_k \left(g_k L b_k ^\dagger+g_k^* L^\dagger b_k \right)+\sum_k \omega_k b_k ^\dagger b_k,
\end{equation}
where $L$ is the coupling operator called Lindblad operator, $b_k ^\dagger \; (b_k)$ is the $k$th mode creation (annihilation) operators of the bosonic bath with frequency $\omega_k$ . Here $g_k$ denotes the coupling strength between the system and the bosonic bath. The pure state of the total system $|\Psi(t)\rangle$ is governed by the standard Schr\"odinger equation, and the reduced density operator $\rho$ for the system of interest only is obtained by tracing out the environmental degrees of freedom $\rho(t)={\rm Tr}_{\rm env} [|\Psi(t)\rangle\langle \Psi(t)|]$. Due to the coupling between the system and its environment, the reduced state of the open system is typically a mixed state.  Several attempts have been made to calculate the geometric phase associated with this non-unitary evolution process including state purification~\cite{Tong2004} and quantum jump approach~\cite{Carollo2003}.  Notably, when the environment memory can be ignored, the quantum jump approach takes on the problem by calculating the phase of Markov stochastic quantum trajectories $|\psi_j \rangle$ for the reduced density operator $\rho=\sum_j p_j|\psi_j \rangle\langle \psi_j|$, and the state purification approach generalizes on the parallel transport condition to find a geometric phase associated with an enlarged system whose partial trace reproduces the reduced density operator for the system of interest. For a generic quantum open system described by (\ref{total_H}),  however,  the master equation governing the reduced density operator may be not approximated by a Markov equation, and the general non-Markovian master equations are not available.  An important feature of our approach is that the geometric phase is directly associated with a single physical process for an arbitrary open system depicted in \eqref{total_H} regardless of the existence of the exact or approximate master equations. Moreover, the quantum trajectory based quantum geometric phase may offer an effective feedback control mechanism that gives rise to robust realizations of the geometric quantum information processing.

The non-Markovian QSD is obtained from Eq.~\eqref{total_H} through specifying the bath state by a set of complex numbers $\{z_k\}$ labeling the Bargmann coherent state of all bath modes and projecting the quantum state $|\Psi_{\rm{tot}}(t)\rangle$ of the total system into the bath state $|z \rangle\equiv \prod_k|z_k \rangle$, we have $|\psi_{z^*}(t) \rangle=\langle z |\Psi_{\rm{tot}}(t)\rangle$, which is called a quantum trajectory and obeys a linear QSD equation~\cite{Diosi1998}
\begin{equation}
    \frac{\partial}{\partial t}|\psi_{z^*}(t) \rangle= \left[-iH_{\rm sys}+Lz_t^*-L ^\dagger \bar{O}(t,z^*)\right]|\psi_{z^*}(t) \rangle,\label{eq_qsd}
\end{equation}
where $z_t^*\equiv -i\sum_k g_k^* z_k^* e^{i \omega_k t}$, $O$ is an operator defined by the functional derivative $\frac{\delta}{\delta z_s^*}|\psi_{z^*}(t) \rangle=O(t,s,z^*)|\psi_{z^*}(t) \rangle$ and $\bar{O}(t,z^*)=\int_0^t \alpha(t,s) O(t,s,z^*)ds$.  Note that the effect of the bath is characterized by the correlation function $\alpha(t,s)=\sum_k |g_k|^2 e^{-i \omega_k (t-s)}$. 
One can interpret $z_k$ as a Gaussian random variable and $z_t^*$ becomes a random process with its statistical mean given by the bath correlation function, i.e., $\mathcal{M}[z_t z_s^*]=\alpha(t,s)$, where $\mathcal{M}[\mathcal{F}]=\frac{1}{\pi}\int d^2z e^{-|z|^2}\mathcal{F}$ represents the ensemble average over noise.

A normalized quantum trajectory is defined by  $|\tilde\psi_{z^*}(t) \rangle=|\psi_{z^*} (t)\rangle/\sqrt{ \langle \psi_{z}(t)|\psi_{z^*}(t) \rangle}$. To calculate the geometric phase associated with this trajectory, we may discretize the wave function as $|\psi_j\rangle=|\tilde\psi_{z^*} (j \Delta t) \rangle$, where $j=0,1,\ldots N$ and $\Delta t=t/N$ is the time step length. For this chain of pure states, the geometric phase $\gamma_G$ is given by the well-known Pancharatnam formula~\cite{Carollo2003,Pancharatnam1956},
\begin{equation}
    \gamma_G=-\arg \left[\langle \psi_0|\psi_1 \rangle \langle \psi_1|\psi_2 \rangle \ldots \langle \psi_{N-1}|\psi_N \rangle \langle \psi_N|\psi_0 \rangle\right],\label{pgp_def}
\end{equation}
where the last term $-\arg[\langle \psi_N|\psi_0 \rangle]$ represents the total phase $\gamma_{\rm tot}$ and the rest constitutes the dynamical phase $\gamma_{\rm dyn}$. It can be readily shown that this definition coincides with the definition using the fiber bundle reference section~\cite{Pati1995} through $|\chi(t) \rangle=\xi(t) |\tilde\psi_{z^*}(t) \rangle$, where $\xi(t)=\langle \tilde\psi_{z}(t)|\tilde\psi_{z^*}(0) \rangle/|\langle \tilde\psi_{z}(t)|\tilde\psi_{z^*}(0) \rangle|$, and the geometric phase is given by 
\begin{align}
    \gamma_G&=i\int \langle \chi(t)|\partial_t|\chi(t) \rangle dt \nonumber\\
    &= i  \int  \dot{\xi}(t)\xi^*(t) dt + i\langle \tilde\psi_z(t)|\partial_t|\tilde\psi_{z^*}(t) \rangle dt \label{def_pati},
\end{align}
where $\partial_t$ represents the derivative with respect to time, and the first term is the total phase, and the second term is the dynamical phase (up to a minus sign), and the integral runs along the path traced by the quantum state for the open system. We may further close the path by the geodesic that connects the two ends of the path, and then convert the line integral to a surface integral. Obviously,  one can always compute the geometric phase by using the trajectories generated from the linear QSD with Eq.~\eqref{pgp_def}  which greatly simplify the calculations.

The ensemble average for the trajectories may be obtained by using the Bargmann coherent basis $\frac{1}{\pi}\int d^2z e^{-|z|^2}|z \rangle\langle z|=\mathbb{1}$. Accordingly, the reduced density operator of the system is given by $\rho(t)=\tr_\mathcal{E} \left[\mathbb{1}\cdot |\Psi_{\rm tot}(t) \rangle\langle \Psi_{\rm tot}(t)|\right]=\mathcal{M}\left[|\psi_{z^*}(t) \rangle \langle \psi_z(t)| \right]$. Next, we show that the geometric phase for the open system can be obtained by the ensemble average based on Eq.~\eqref{pgp_def}.
To calculate the average dynamical phase $\bar \gamma_{\rm dyn}$, we first note that $\langle \psi_z[j\Delta t]|\psi_{z^*}[(j+1) \Delta t] \rangle=\langle \psi_z(j\Delta t)|\psi_{z^*}(j \Delta t) \rangle+dt\langle \psi_z(j\Delta t)|\partial_t|\psi_{z^*}(j \Delta t) \rangle$. Using Eq.~\eqref{eq_qsd} and taking the continuous time limit $\Delta t \rightarrow 0$, $\bar \gamma_{\rm dyn}$ is given by
\begin{align}
    \bar \gamma_{\rm dyn}&=-i \int_0^t d\tau \mathcal{M}\left[\langle \psi_z(\tau)|\partial_t|\psi_{z^*}(\tau) \rangle \right] \nonumber\\
    &= -i\int_0^t d\tau \mathcal{M}\left[\langle \psi_z(\tau)|h(\tau)|\psi_{z^*}(\tau) \rangle \right],
\end{align}
where $h(\tau)=-iH_{\rm sys}+Lz_\tau^*-L ^\dagger \bar{O}(\tau,z^*)$, and we have used the relation $\mathcal{M}[\langle \psi_z(t)|\psi_{z^*}(t) \rangle]=\tr[\rho(t)]=1$, which also implies $\mathcal{M}[\langle \psi_z(t)|\partial_t|\psi_{z^*}(t) \rangle]$ is pure imaginary. The total average phase is simply $\bar \gamma_{\rm tot}=\arg \left[\mathcal{M}\left(\langle \psi_z(0)|\psi_{z^*}(t) \rangle\right) \right]$. This total phase for the non-cyclic evolution may be measured by an interferometer~\cite{interf}, where the input (initial) state goes through one arm and $|\psi_{z^*}(t) \rangle$ goes through another. The pure state $|\psi_{z^*}(t) \rangle$ can be obtained by a measurement on the bath at time $t$, with an outcome labeled $z^*$~\cite{strunz_qsdm}.

The ensemble average for the dynamical phase may be carried out with Novikov's theorem~\cite{Diosi1998} $\mathcal{M}[P_t z_t]=\int_0^t ds \mathcal{M}[z_t z_s^*]\mathcal{M}[\frac{\delta P_t}{\delta z_s^*}]=\int_0^t ds \alpha(t,s)\mathcal{M}[\frac{\delta P_t}{\delta z_s^*}]$. The average geometric phase is then given by $\bar \gamma_G=\bar \gamma_{\rm tot}-\bar \gamma_{\rm dyn}$. It's worth pointing out that in our definition of the geometric phase, the form of the system Hamiltonian $H_{\rm sys}$ and the system-bath coupling operator $L$ are entirely general. Moreover, no assumptions including weak coupling or Markov approximations have been assumed. Hence, the formulation of the geometric phase has incorporated all the information about the environment without using an explicit master equation. Interestingly, when the $\bar{O}$ operator itself is noise independent, we have
\begin{equation}
    \bar \gamma_{\rm dyn}=-\int_0^t d\tau \tr \left[H_{\rm sys} \rho(\tau) \right] +2\mathrm{Im} \tr \left[L ^\dagger \bar{O}(\tau) \rho(\tau)\right],\label{gdyn_noz}
\end{equation}
where $\rho(\tau)$ is the reduced density operator of the system at time $\tau$. Again, using the Novikov's theorem~\cite{Yuetal1999}, this dynamical phase can still be expressed in terms of the density operator for the system when the $\bar{O}$ operator is noise-dependent following a similar approach in the derivation of master equations from QSD~\cite{qsd_meq}. Since both the total and dynamical phases are in principle measurable, our definition represents an \emph{operational} formulation of the geometric phase under general non-Markovian open system dynamics. It is also worth pointing out that this definition is independent of the unraveling chosen (see Appendix A for details) and one is free to choose other types of projection~\cite{strunz_qsdm} than the coherent state used here.

\section{Examples}
\begin{figure}[t!]
    \includegraphics[scale=.6]{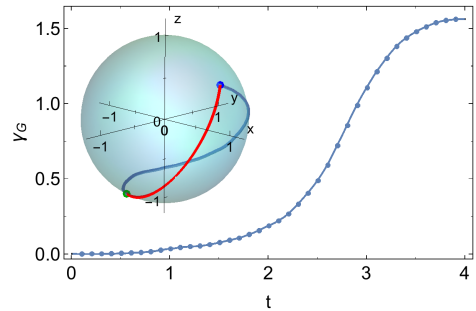}\\
    \caption{The geometric phase $\gamma_G$ for a single trajectory in the dissipative model as a function of time $t$, with $\omega=\lambda=1$ and $\theta=1$. The solid line is calculated from the Pantcharatnam formula Eq.~\eqref{pgp_def} and the dots are calculated as half the solid angle enclosed by the closed path formed by the QSD trajectory whose ends are joined by a geodesic. Inset: Bloch sphere representation of the normalized path generated by the QSD (blue) whose ends are joined by a geodesic (red) to form a closed path.}\label{rwa_1trj}
\end{figure}

\begin{figure}[t!]
    \includegraphics[scale=.6]{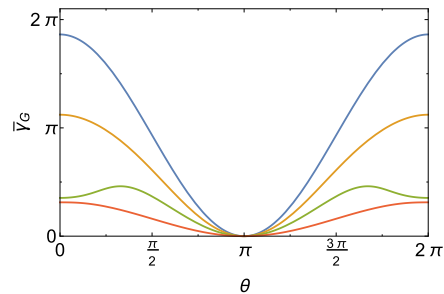}\\
    \caption{The geometric phase $\bar \gamma_G$ for the dissipative model as a function of the  initial state parameter $\theta$ for different $\gamma$'s: from top to bottom, $\gamma=0.1,\; 0.5,\; 1.2,\; 100$, with $\lambda=1$, $\Gamma=1$, $\Omega=0$ and $t=2\pi/\omega$.}
    \label{rwagtha}
\end{figure}

 The general definition of the  geometric phase of a non-Markovian open system will be illustrated by two physically interesting models that will allow the explicit analytical solutions. 
 More general case will inevitably require numerical simulations, which prove to be particularly efficient when the quantum trajectories are employed \cite{Yuetal1999,nqubit,Jing2010}. The models to be considered are the dissipative spin-boson model and dephasing two-level system. In each case, the general non-Markovian geometric phase is explicitly calculated, giving rise to an important geometric phase that is not found in the case of Markov limit.  To be more specific, we prepare the two-level system in a pure state characterized by a Bloch vector $(\sin \theta,0,\cos \theta)$. For the dissipative two-level system described  by the system Hamiltonian $H_{\rm sys}=\omega \sigma_z/2$ and the coupling operator $L=\lambda \sigma_-$,   the exact solution may be obtained~\cite{Diosi1998} (see Appendix B for more details), and the $\bar{O}$ operator is explicitly given by $\bar{O}(t)=F(t)\sigma_-$  where the function $F(t)$ can be analytically obtained.  Note that here, although our formulation is valid for an arbitrary correlation function,  however, for simplicity,  we have assumed the spectrum of the bath to be of the Lorentzian form, and the correlation function is accordingly given by $\alpha(t,s)=\Gamma \gamma \exp[-\gamma|t-s|-i \Omega (t-s)]/2$ at zero temperature. This spectrum is particularly convenient for studying non-Markovian/Markov transition as $\gamma \rightarrow \infty$ corresponds to the Markov limit. Throughout this example, we set $\Gamma=1, \Omega=0$.  By definition, the geometric phase acquired by the quantum trajectory will be dependent on the path transversed by the state only. For the two-level system it should be equal to half the solid angle enclosed by the closed path formed by joining the open ends with a geodesic. Here, we plot one realization of the non-Markovian QSD trajectory with its associated geometric phase in Fig.~\ref{rwa_1trj} using Eq.~\eqref{pgp_def}, where we see that the geometric phase exactly coincides with half the solid angle enclosed. Since this model is analytically solvable, we have an exact and analytical expression for the geometric phase (see Eq.~\eqapxrwa{} in Appendix B). When $\theta=\pi$, the wave function for the total system becomes $|\Psi(0)\rangle=|\downarrow \rangle\otimes \prod_k |0 \rangle_k$, which is an eigenstate of the total system. Since the total Hamiltonian is time-independent in this case, this state acquires zero geometric phase, as expected. At $t=2\pi/\omega$ in the Markov limit, the expression for $\bar \gamma_G$ simplifies to
\begin{align*}
    \frac{\omega  \left[\cos \theta +1\right]  \left(1-e^{-\frac{2 \pi  \lambda^2 }{\omega }}\right)}{2 \lambda^2 }.
\end{align*}
Further taking the weak-coupling limit, we have $\bar\gamma_G\approx \gamma_G^{(M)}=\pi\left[\cos \theta +1\right] $, which reduces to the geometric phase of an isolated system. A remarkable feature of this model is that when the system is interacting with a memory-less bath, the maximally attainable geometric phase is reduced by a factor of (taking $\omega=1$)
\begin{align}
    \frac{1-e^{-2 \pi  \lambda^2 }}{2 \pi  \lambda^2 }.
\end{align}
Therefore, as the coupling strength ratio $\lambda$ increase,  the range of possible geometric phase acquired by the state is reduced, which is consistent with the weak-coupling limit. 
On the other hand, the environmental memory can increase the controllable range of attainable geometric phase. To see this, we analytically calculated the geometric phase in the small $\gamma$ limit, i.e. strong bath memory regime, and the geometric phase is given by
\begin{align*}
    \bar\gamma_G= \gamma_G^{(M)} \left[1-\frac{\gamma  \Gamma \lambda^2 }{2}\right],
\end{align*}
Therefore, the system-environment may be engineered to give rise to a correlation parameter $\gamma$ that will generate a \emph{memory-assisted}  geometric phase, where the range of attainable phase can be improved with the help of environmental memory. Using Eq.~\eqapxrwa{} in the Appendix B, we display an example in Fig.~\ref{rwagtha}, where the red line corresponds to the memory-less case where the range of attainable geometric phase is greatly reduced, and the blue curve corresponds to the case where the full range $[0,2\pi]$ is almost recovered with the help of environment memory effect. 
The example has clearly demonstrated that the environmental memory can fundamentally change the geometric phase for the open quantum systems. Our analysis shows that when quantum systems are engaged by the noises with memory,  the geometric phase is not predicted by the  well-known Markov behavior that is familiar from the quantum theory of Markov open systems or closed systems.

\begin{figure}[t!]
    \includegraphics[scale=.6]{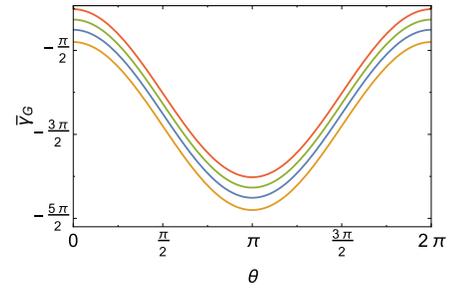}
    \caption{The geometric phase $\bar \gamma_G$ for the dephasing model as a function of the  initial state parameter $\theta$ for different $\gamma$'s: from top to bottom, $\gamma=100,\; 7,\; 0.3,\; 0.7$, with $\lambda=1$, $\Gamma=1$, $\Omega=\omega$ and $t=2\pi/\omega$.}
    \label{dpgtha}
\end{figure}

Our exact analysis on the non-Markovian geometric phase can also be extended to the pure dephasing model.  By employing this simple quantum open system model, we can clearly show how the environmental memory affects the onset of the geometric phase.  The total Hamiltonian (\ref{total_H}) takes the following form: The system Hamiltonian of the model is given by $H_{\rm sys}=\omega \sigma_z/2$ and the coupling to the bath is described by $L=\lambda \sigma_z$. The $\bar{O}$ operator of this model is exactly given by $\bar{O}(t)=\lambda\int_0^t ds \alpha(t,s) ds \sigma_z$. We can analytically solve the QSD equation and derive the corresponding geometric phase after taking the average. Remarkably, when $\Omega=0$, the geometric phase at time $t=2\pi/\omega$ is $\bar \gamma_G=\pi \left(\cos \theta-1\right)$, which the same as a closed system $H=\omega\sigma_z/2$, indicating that the geometric phase in this case is robust against dephasing. This agrees with the geometric phase obtained in the Markov  case via quantum jumps~\cite{Carollo2003}, since the dynamics is actually Markov when $\Omega=0$. When $\Omega\neq 0$, the geometric phase is a complex but analytical function (see Eq.~\eqapxsz{} in Appendix C). We can expand the geometric phase in powers of $1/\gamma$ as
\begin{align*}
    \bar \gamma_G=\gamma_G^{(M)}-\frac{\pi  \Gamma  \lambda ^2 \Omega }{\gamma  \omega }+\frac{\Gamma  \lambda ^2 \Omega }{\gamma ^2}+O \left[\frac{1}{\gamma^3}\right],
\end{align*}
where $\gamma_G^{(M)}=\pi \left(\cos \theta-1\right)$ is the geometric phase in the Markov limit, which also coincides with the closed system case. Taking $\Omega=\omega$ and $t=2\pi/\omega$, the geometric phase can be simplified as

\begin{align}
    \bar \gamma_G=\gamma_G^{(M)}-\frac{\gamma  \Gamma  \lambda ^2 \left[\pi  \left(\gamma ^2+\omega ^2\right)+\gamma  \omega  \left(e^{-\frac{2 \pi  \gamma }{\omega }}-1\right)\right]}{\left(\gamma ^2+\omega ^2\right)^2},\label{eq_gp_rwa_res_simp}
\end{align}
i.e. the geometric phase is subject to a shift which is \emph{independent} of the initial state parameter $\theta$, and the range of attainable geometric phase is unaffected. This shift is maximized when $\gamma \approx \Omega$, at about $1.32$.

Using Eq.~\eqref{eq_gp_rwa_res_simp}, we plot the exact geometric phase $\bar \gamma_G$ for the dephasing model as a function of the bath memory parameter $\gamma=1/\tau$ ($\tau$ is the memory time) and initial state parameter $\theta$ with $\lambda=1$, $\Gamma=1$, $\Omega=\omega$ and $t=2\pi/\omega$ in Fig.~\ref{dpgtha}. As $\gamma$ increases, the geometric phase first decreases then increases asymptotically to the Markov limit.

We emphasize again that the non-Markovian quantum phase can always be calculated from the general QSD equation without assuming the exact solvability, and the geometric phase defined here can be used whenever the QSD approach is applicable, for example with finite temperature bath \cite{qsd_fnt}, fermionic bath \cite{qsd_ferm} and other types of bath correlation functions.~\cite{qsd_gen}. In this case, the existing perturbation technique~\cite{Yuetal1999} or numerical algorithms for the QSD equation~\cite{Li2014Luo2015} can be readily employed, such that the geometric phase can still be obtained under general non-Markovian open system dynamics. Specifically,  the geometric phase for a bosonic bath may be calculated with
\begin{align}
    &\mathcal{I}=\mathcal{M}\left[\langle \psi_z(0)|\psi_{z^*}(\Delta t) \rangle \right]\times\mathcal{M}\left[\langle \psi_z(\Delta t)|\psi_{z^*}(2 \Delta t) \rangle \right] \nonumber\\
    &\times \ldots \times \mathcal{M}\left[\langle \psi_z(T-\Delta t)|\psi_{z^*}(T) \rangle \right]\times \mathcal{M}\left[\langle \psi_z(T)|\psi_{z^*}(0) \rangle \right], \nonumber\\
    &\bar \gamma_G=-\arg(\mathcal{I}),
\end{align}
where $T$ is the final evolution time under consideration. We have verified the results using the generic algorithm in~\cite{Li2014Luo2015} and found a perfect agreement between the numerical result and the analytical results presented above.

\section{Conclusion}

In this paper we introduced a generic formulation of non-Markovian geometric phase for quantum open systems by means of a set of non-Markovian quantum trajectories. This approach represents a systematic and computable way to study the geometric phase in general non-Markovian open systems without relying on the existence of a master equation or purification while at the same time fully accounting for the memory effect of the bath. Additionally, this approach can give a hierarchical approximation of the geometric phase with respect to the degree of bath memory effects, starting with a Markov case, followed by a post-Markov analysis~\cite{Yuetal1999}, all the way up a full non-Markovian treatment. This generic formalism of the geometric phase places no restriction on the system Hamiltonian, the system-bath coupling operator $L$ or the bath correlation function, ensuring its wide application domain. By examining the temporal behavior of quantum trajectories 
under the influence of the environments with memories, we demonstrated the onset of a new geometric phase of open quantum systems not previously encountered in the Markov limit or in the 
case of closed quantum systems. This new geometric phase will arise, for example, in a quantum information processing device, where the environment cannot be approximated by a Markov noise
such as in the case of structured medium or strong coupling regimes. As demonstrated in this paper,  the environmental memory can produce a wide adjustable range for the geometric phase, which is of fundamental importance in realizing universal geometric quantum computation. We also showed that in any implementation of the geometric phase for quantum information processing, the environmental effects must be properly taken into account to maximize the accuracy of the quantum gate operations.

\begin{acknowledgments}
This work is supported by NSF PHY-0925174.  J. Q. You is supported by the National Natural Science Foundation of China No.~91421102 and the National Basic Research Program of China No.~2014CB921401. L. Wu is supported by the Spanish MICINN (No. FIS2012-36673-C03-64203 and Grant No. FIS2015-67161-P). H.Q. Lin, J.Q. You and D.W. Luo also acknowledge support from NSAF U1530401.
\end{acknowledgments}

\onecolumngrid
\appendix

\section{Geometric nature for the ensemble average geometric phase}
%
Here we show that the ensemble average $\bar \gamma_G$ is also geometric in nature. Consider the composite total wave function for the system and bath $|\Psi(t) \rangle$, which we can also discretize as $|\Psi_j\rangle=|\Psi (j \Delta t) \rangle$, where $j=0,1,\ldots N$ and $\Delta t=t/N$ is the time step length. Since the total wave function describing the system plus bath state remains pure during evolution, we can also use the standard Pancharatnam approach to ensure the geometric nature for geometric phase associated this composite pure state wave function. Explicitly, the ensemble average geometric phase can be written as
\begin{equation*}
    \bar \gamma_G=-\arg \left[\langle \Psi_0|\Psi_1 \rangle \langle \Psi_1|\Psi_2 \rangle \ldots \langle \Psi_{N-1}|\Psi_N \rangle \langle \Psi_N|\Psi_0 \rangle\right].
\end{equation*}
To calculate each of the inner product above, we insert the identity operator $\mathbb{1}=\frac{1}{\pi}\int d^2z e^{-|z|^2}|z \rangle\langle z|$. Under the QSD approach, the inner product can be written in terms of the ensemble average of the trajectories since
\begin{align*}
    \langle \Psi_i|\Psi_j \rangle &= \frac{1}{\pi}\int d^2z e^{-|z|^2} \langle \Psi_i|z \rangle\langle z|\Psi_j \rangle\\
    &=\mathcal{M} \left[\langle \psi_z(i \Delta t)|\psi_{z^*}(j \Delta t) \rangle\right]=\mathcal{M} \left[\langle \psi_i|\psi_j \rangle\right].
\end{align*}
This gives us a possible way to carry out the ensemble average for the geometric phase in non-Markovian open systems that's also a geometric entity. Since we only insert a identity operator above, this way to calculate the inner product is also independent of how we write the identity operator (for example, other than using coherent states), which corresponds to other types of open system unraveling.

\section{Calculation details for the geometric phase in the dissipative spin-boson model}

For the first example in the main text described by the system Hamiltonian $H_{\rm sys}=\omega \sigma_z/2$ and the coupling operator $L=\lambda \sigma_-$, the $O$-operator does not contain noise terms and can be exactly given~\cite{Diosi1998}. Denote
\begin{align}
    O(t,s)=f(t,s)\sigma_-,
\end{align}
we have
\begin{align}
    \partial_t f(t,s)= \left[i \omega+\lambda F(t)\right] f(t,s),
\end{align}
where $F(t)=\int_0^t ds \alpha(t,s)f(t,s)$. For simplicity, we have assumed the spectrum of the bath to be Lorentzian, and the correlation function is accordingly given by $\alpha(t,s)=\Gamma \gamma \exp[-\gamma|t-s|-i \Omega (t-s)]/2$ at zero temperature. In this case, $F(t)$ may be analytically obtained,
\begin{align}
    F(t)=\frac{\gamma -\Omega _p \tan \left[-\frac{\Omega _p t}{2}+\tan ^{-1}\left(\frac{\gamma -i (\omega -\Omega )}{\Omega _p}\right)\right]-i (\omega -\Omega )}{2 \lambda },
\end{align}
where 
\[
\Omega_p=\sqrt{-\gamma ^2+2 \gamma  \left[\lambda ^2+i (\omega -\Omega )\right]+(\omega -\Omega )^2}.
\]

Using the ensemble average of Eq.~(5), we arrive at the analytical form for the geometric phase in the dissipative spin-boson model $\bar \gamma_g=\bar \gamma_{\rm tot}-\bar \gamma_{\rm dyn}$, where
\begin{align}
    \bar \gamma_{\rm tot}&=-\arg \left\{e^{-i \omega t/2}\left[1-\cos \theta +\exp[i \omega t-\lambda g^*(t)](1+\cos \theta)\right]\right\},\nonumber\\
    \bar \gamma_{\rm dyn}&=\int_0^t \frac{\omega}{2}-\exp[-2 \lambda \mathrm{Re}\{g(s)\}] \cos^2(\theta/2)\left[\omega+2 \lambda \mathrm{Im} \{F(s)\} \right]ds,
\end{align}
and $g(t)=\int_0^t F(s)ds$.

\section{Calculation details for the geometric phase in the pure dephasing model}
The second example in the main text also admits an exact analytical solution. Assuming an Lorentzian spectrum for the bath at zero temperature as example 1. Using the ensemble average of Eq.~(5), the geometric phase for the non-Markovian pure dephasing model is given by
\begin{align}
    \bar \gamma_{g}&=\arg \left[\left(\cos \theta +1-(\cos \theta -1) e^{i  \omega t}\right) \exp \left(-\frac{\gamma  \Gamma  \lambda ^2 \left((\gamma +i \Omega )t+e^{-(\gamma +i \Omega ) t }-1\right)}{2 (\gamma +i \Omega )^2}-\frac{i  \omega t }{2}\right)\right] \nonumber\\
    &-\frac{e^{-\gamma  t} \left[2 \gamma  \Gamma  \lambda ^2 \left(\Omega  e^{\gamma  t} \left(\gamma  (\gamma  t-2)+ \Omega ^2 t\right)+(\gamma^2 -\Omega^2 ) \sin  \Omega t+2 \gamma  \Omega  \cos  \Omega t\right)- \omega t \left(\gamma ^2+\Omega ^2\right)^2 \cos \theta e^{\gamma  t}\right]}{2 \left(\gamma ^2+\Omega ^2\right)^2}.
\end{align}
At $t=2\pi/\omega$, we have
\begin{align}
    \bar \gamma_g=&\frac{e^{-\frac{2 \pi  \gamma }{\omega }}}{2 \omega  \left(\gamma ^2+\Omega ^2\right)^2}\times \left[2 e^{\frac{2 \pi  \gamma }{\omega }} \left(-\pi  \left(\gamma ^2+\Omega ^2\right) \left(\gamma ^2 \omega +\gamma  \Gamma  \lambda ^2 \Omega +\omega  \Omega ^2\right)+\gamma ^2 \Gamma  \lambda ^2 \omega  \Omega +\pi  \omega  \left(\gamma ^2+\Omega ^2\right)^2 \cos \theta\right) \right.\nonumber \\
    &\left.+\gamma  \Gamma  \lambda ^2 \omega  \left(\left(\Omega ^2-\gamma ^2\right) \sin \frac{2 \pi  \Omega }{\omega }-2 \gamma  \Omega  \cos \frac{2 \pi  \Omega }{\omega }\right)\right]
\end{align}

With $\Omega=\omega$, it simplifies to Eq.~(9) in the main text.
\twocolumngrid




\end{document}